\documentclass[doublecol,linenumbers,figures]{epl2} 

\usepackage{hyperref,verbatim}
\usepackage{dcolumn,epstopdf,color,amsmath}
\newcommand{\RE}[1]{\operatorname{Re}\left \{#1\right\}}
\newcommand{\IM}[1]{\operatorname{Im}\left\{ #1\right\}}

\newcommand{\oab}[1]{\omega_{#1}}
\newcommand{\Oab}[1]{\Omega_{#1}}

\newcommand{\dab}{\delta_{AB}}

\newcommand{\kv}{\mathbf k}

\title{Fano-like anti-resonances in strongly coupled binary Coulomb systems}
\shorttitle{} 

\author{L. Silvestri\inst{1}, G. J. Kalman\inst{1}, Z. Donk\'o\inst{1,2}, P. Hartmann\inst{1,2}, H. K\"{a}hlert\inst{1,3}}
\shortauthor{L. Silvestri \etal}

\institute{                    
  \inst{1} Department of Physics, Boston College, 140 Comm. Ave, Chestnut Hill, MA 02467, USA\\
  \inst{2} Institute for Solid State Physics and Optics, Wigner Research Centre for Physics, Hungarian Academy of Sciences, H-1121 Budapest, Konkoly-Thege Mikl\'os str. 29-33, Hungary\\
  \inst{3} Institut f\"ur Theoretische Physik und Astrophysik, Leibnizstr. 15, 24098 Kiel, Germany
}

\abstract{
Molecular Dynamics (MD) simulations of a strongly coupled binary ionic mixture have revealed the appearance of sharp minima in the species resolved dynamical density fluctuation spectra. This phenomenon is reminiscent of the well-known Fano anti-resonance, occurring in various physical processes. We give a theoretical analysis using the Quasi Localized Charge Approximation, and demonstrate that the observed phenomenon in the equilibrium spectrum is a novel manifestation of the Fano mechanism, that occurs at characteristic frequencies of the system different from the conventional classical Fano frequencies.}

\begin{document}

\maketitle


The pioneering work of Ugo Fano  \cite{Fano1935,Fano1961} on asymmetric profiles of absorption spectra of noble gases has reverberated in many diverse fields of physics. The ubiquity of Fano resonances is mostly due to the simple physical foundation of the phenomenon: interference between two quantum states. Recently it has been pointed out that the Fano resonance effect can be understood on the basis of a classical model \cite{Joe2006} and therefore it is a phenomenon that should occur in classical systems as well. Indeed, such an occurrence has been demonstrated in metamaterials and plasmonic nanostructures \cite{Papasimakis2009,Miroshnichenko2010,Luk'yanchuk2010}. Much of this research is now motivated by the possible applications of these works to areas such as electromagnetically induced transparency \cite{Papasimakis2009} and polarization spectro-tomography \cite{Shafiei2013}. 

It is well understood that the physical explanation of the Fano effect is to be sought in the behavior of the response of a system of two coupled damped harmonic oscillators to an input signal. In particular, the response relating the output signal of one of the oscillators to the input signal directed at the same oscillator is affected by the presence of the other ``spectator'' oscillator: the oscillation amplitude of the driven oscillator vanishes at the shifted (by the coupling) natural frequency of the second oscillator. At this frequency the first oscillator experiences two forces, one due to the driving force (out of phase) and the other from the second oscillator (in phase), which cancel each other. The result is the appearance of sharp minima (anti-resonances) in the frequency spectrum of the real (non-dissipative) part of the response. While most of the attention in relation to the Fano resonance has focused on this real part of the response, it is clear that the imaginary (dissipative) part must be affected as well, as corroborated by recent experiments. What, however, is not widely recognized is that there must be a corollary to the Fano phenomenon as demanded by the Fluctuation-Dissipation Theorem (FDT): in addition to the response function, the \textit{equilibrium fluctuation spectrum} of a strongly coupled binary system has to display a similar spectral feature \cite{Liu2009}. However, the details of this process are not a priori obvious. What we show below is that in fact the equilibrium partial density fluctuation spectra also develop sharp minima at characteristic frequencies which are, however, quite different from the Fano resonance frequencies and may be regarded as novel natural frequencies of the system. In this Letter we report the observation by Molecular Dynamics (MD) simulation of this novel phenomenon and provide a theoretical explanation for it.

The system we have studied is a strongly coupled binary ionic mixture (BIM), consisting of two species of positively charged particles with charges, $Z_A$, masses, $m_A$, and densities, $n_A$, that interact via a Coulomb potential. A rigid neutralizing background of negatively charged particles assures overall charge neutrality. This system is characterized by the asymmetry parameters $Z = Z_2/Z_1$, $m = m_2/m_1$, $\rho = n_2/n_1$, $N_A$ being the number of particles of species A, and the coupling constant  $\Gamma$ defined with reference to species 1: $\Gamma = (Z_1e)^2/a_1 k_BT $. The Wigner-Seitz radii are defined as $a_A = (3/4\pi n_A)^{1/3}$, and $a = \sqrt{a_1a_2}$. 

The MD simulations on the system are performed both in its strongly coupled liquid and solid phases. Our code is based on the Particle-Particle Particle-Mesh algorithm proposed and described in details by Hockney and Eastwood \cite{PPPM}. We trace 10\,000 point-like particles, with random initial positions, in the liquid phase simulations and 8192 point-like particles, initialized in a bcc configuration, for the solid phase simulations within a cubic simulation box (with side length $L$). During the course of the simulation, data for the microscopic densities and currents of the two particle species are Fourier analyzed, to yield the partial dynamical density-density correlation functions (pDCF), $S_{AB}(\kv,\omega)$, and the species resolved (partial) longitudinal current-current correlation functions (pCCF), $L_{AB}(\kv,\omega)$ \cite{Hansen1975}.

\begin{figure}[htb]
\onefigure[width=.9\columnwidth]{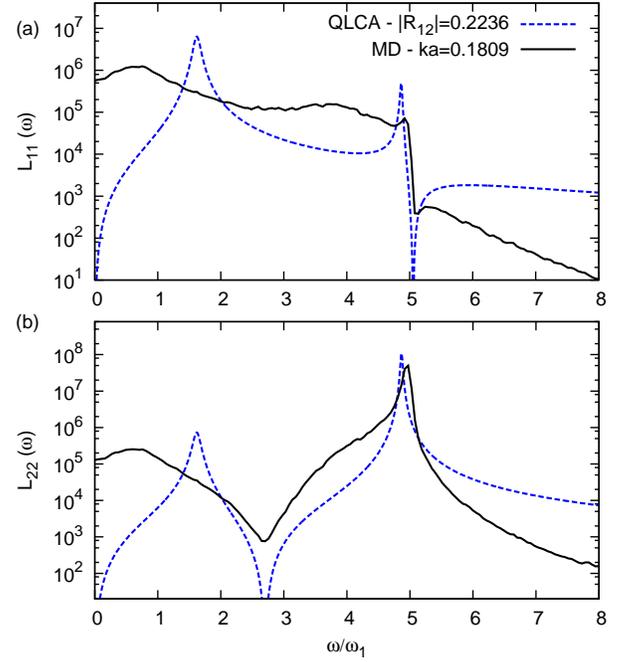}
\caption{(Color online) MD (solid black line) and extended QLCA results (dashed blue line) for a) $L_{11}$, b) $L_{22}$ in the liquid phase, $\Gamma=200$, $Z=0.7$, $m=0.2$, $c=0.1$.}
\label{liquid}
\end{figure}

\begin{figure}[htb]
\onefigure[width=.9\columnwidth]{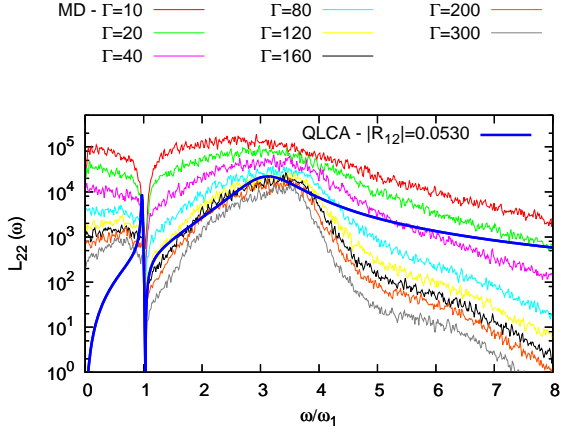}
\caption{(Color online) $L_{22}$ at $ka = 0.1809$ for a wide range of $\Gamma$ values together with the extended QLCA (solid blue line) calculated at $\Gamma = 300$, $Z=0.5$, $m=0.02$, $c=0.9$.}
\label{gammarange}
\end{figure}

\begin{figure}[htb]
\onefigure[width=.9\columnwidth]{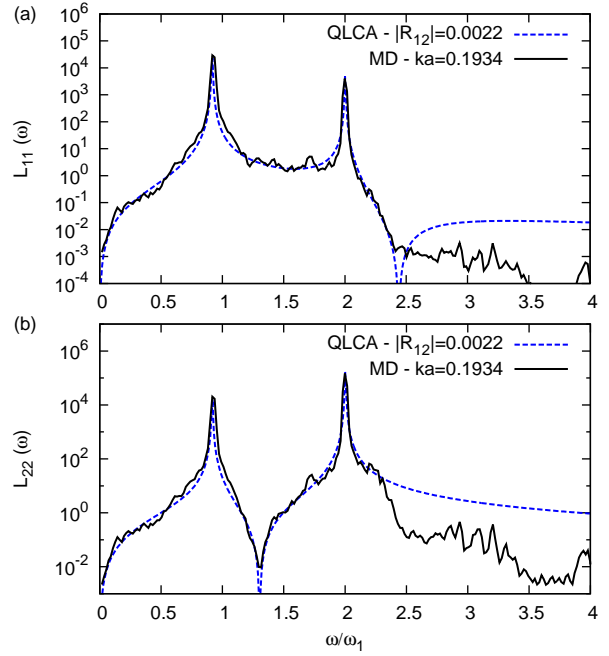}
\caption{(Color online) Same as in Fig.~\ref{liquid} for $\Gamma=10\,000$, $Z=0.7$, $m=0.2$, $c=0.5$.}
\label{BCC}
\end{figure}

The resulting spectra with parameters $Z=0.7$, $m=0.2$, and at a low concentration of the heavy component, $c=0.1$, for the strongly coupled liquid phase $(\Gamma = 200)$ are shown in Fig.~\ref{liquid}. In Fig.~\ref{gammarange} we show results for $Z=0.5$, $m=0.02$, $c=0.9$, for a series of $\Gamma$ values. Figure \ref{BCC} displays spectra at much higher $\Gamma$ values, when the system is in the solid phase. We simulate the bcc structure, in which the system crystallizes in the case of equal concentrations. In the simulations the $k$ values are multiples of $k_{\rm {min}} = 2\pi /L$, in the figures we show data for $k_{\rm min}$, allowing extrapolation to the $k=0$ limit. The collective modes of the system are identified as peaks in these plots. In addition to these peaks, we observe the rather dramatic formation of deep minima, with a depth of several orders of magnitude, especially in the ($L_{22}$) spectra of the lighter component. We contend that the formation of these minima is a hitherto unexplored consequence of the Fano effect, as it is demonstrated by the theoretical analysis below.

In order to obtain the partial response function, $\chi_{AB}(\kv,\omega)$ \cite{Kalman1984}, we adopt a description based on the Quasi-Localized Charge Approximation (QLCA) \cite{Kalman1990,Golden2000}. The relationship between $S_{AB}(\kv,\omega)$ and $\chi_{AB}(\kv,\omega)$ is established by the FDT \cite{Kubo1957,Kubo1966}
\begin{eqnarray}
S_{AB}(\mathbf k, \omega) &= &- \frac{1}{\beta \pi \omega\sqrt{n_An_B}} \IM{\chi_{AB}(\mathbf k ,\omega)}, \\
L_{AB}(\mathbf k, \omega) &=& - \frac {\omega}{k^2}\frac{1}{\beta \pi \sqrt{n_An_B}} \IM{\chi_{AB}(\mathbf k ,\omega)}.
\end{eqnarray}
where $\IM{\chi_{AB}}$ are the elements of the imaginary part of $\chi$.

Physically, $\IM {\chi}$ is linked to the damping in the system and therefore it is crucial to properly account for the dissipation mechanism in the theoretical model. In order to do this, we complement the QLCA formalism by introducing a two-body collisional matrix $\gamma_{ij}^{AB}\left ( \mathbf x^A_i - \mathbf x^B_j \right)$, $\gamma_{ij}^{AB} = \gamma_{ji}^{BA}$, with its Fourier transform $\gamma^{AB}(\kv)$. The elements of $\gamma_{ij}^{AB}$ represent collisions between pairs of particles of the same species (intraspecies) and collisions between pairs of particles of different species, (interspecies). We assume that the effect of collisions manifests itself as a drag force between two particles, proportional to their perturbed relative velocity. Thus, the equation of motion for small longitudinal oscillations, (we are interested only in the behavior of the longitudinal mode here) $\xi_i^A$, around the quasi static equilibria, $\mathbf{x}_i^A$, for the $i$-th particle of species $A$ is
\begin{eqnarray}
m_A \ddot \xi_i^A & -& \sum_B\sum_{j}\gamma_{ij}^{AB}\left( \dot \xi_i^A - \dot \xi^{B}_{j}\right) \nonumber \\
&+&\sum_B\sum_{j}K^{AB}_{ij}\xi_j^B=Z_Ae E(\mathbf{x}_i^A,t),
\label{eom}
\end{eqnarray}
where $K^{AB}_{ij}$ is the (species matrix) element of the force obtained from the harmonic approximation of the potential \cite{Golden2000}, and $E(\mathbf{x}_i^A,t)$ the external longitudinal electric field. The collisional matrix can now be averaged over all the pairs with the aid of the equilibrium pair correlation function and absorbed in a drag matrix $\mathbf R$ with elements 
\begin{eqnarray}
R_{AB}(\kv) &=& -\sqrt{\frac{n_An_B}{m_Am_B}} \left \{ \int d^3r\gamma^{AB}(\mathbf r) \left[ 1 + h_{AB}(r) \right ]e^{-i\mathbf k \cdot \mathbf r} \right. \nonumber \\
& - & \left . \delta_{AB} \sum_C \frac{n_C}{n_A} \int d^3r\gamma^{AC}(\mathbf r) \left[ 1 + h_{AC}(r) \right ] \right \}, \label{dragmatrix}
\end{eqnarray}
where $\mathbf r = \mathbf x^A_i - \mathbf x^B_j$, and $h_{AB}(r)$ is the pair correlation function. The $\mathbf R$ matrix describes the total drag force acting on a selected particle. The first term represents the force generated by the motion of all the other particles in the system, while the second term is due to the velocity of the selected particles itself in the equilibrium environment of all the others. The drag matrix now combined with the longitudinal element of the original QLCA dynamical matrix, given as
\begin{eqnarray}
C_{AB}(\mathbf k) &=& -\frac{\oab{AB}^2}{4\pi} \int d^3r \psi^L(r) \left [ 1 + h_{AB}(r) \right ] e^{-i\mathbf{k\cdot r}} \nonumber \\
& & + \dab \sum_{C} \frac{\Omega_{AC}^2}{4\pi} \int d^3r \psi^L(r) \left [ 1 + h_{AC}(r) \right ]  \nonumber \\
& & + \dab \sum_{C} \frac{1}{3} \Oab{AC}^2, \\
\psi^L (r) &= & \frac{1}{r^3}\left [ 3 \left ( \frac{\kv \cdot \mathbf r}{kr} \right )^2 - 1 \right ] - \frac{4\pi}{3}\delta(\mathbf r), \label{long}
\end{eqnarray}
\begin{equation}
\oab{AB}^2 = \frac{4\pi Z_AZ_Be^2\sqrt{n_A n_B}}{\sqrt{m_A m_B}},\Oab{AB}^2 = \frac{4\pi Z_AZ_Be^2 n_B}{m_A }\label{plasmaEinstein}
\end{equation}
provides the extended dynamical matrix 
\begin{equation}
G_{AB}(\kv,\omega) = i \omega R_{AB}(\kv) + C_{AB}(\kv).
\end{equation}
Eq.~\ref{long} is the longitudinal projection of the Coulomb potential, in the harmonic approximation. Equations~\ref{plasmaEinstein} define the nominal plasma and Einstein frequencies respectively. 

In this Letter we focus on the mode behavior in the $k\rightarrow 0$ long wavelength limit only. In relation to the damping, momentum conservation requires that the intraspecies contributions vanish in this limit: thus we are left with the interspecies collision term only. The components of the $G_{AB}$ matrix at $k=0$ then become
\begin{eqnarray}
G_{11} = i\omega \frac{n_2}{m_1}\bar\gamma + \omega_{11}^2 + \frac{\Omega_{12}^2}{3},\\
G_{22} = i\omega\frac{n_1}{m_2}\bar\gamma + \omega_{22}^2 + \frac{\Omega_{21}^2}{3},\\
G_{12} = - i \omega\sqrt{\frac{n_1n_2}{m_1m_2}}\bar \gamma + \frac{2}{3}\omega_{12}^2,
\label{qlcak0}
\end{eqnarray}
where $\bar\gamma = \int d^3r \gamma^{12}[1 + h_{12}(r)]$. This latter depends on the system parameters, in a largely unknown fashion, but for the purpose of this work that dependence is irrelevant, and we treat $\bar \gamma$ as an adjustable parameter. Finally, the longitudinal part of the partial response function matrix is obtained as 
\begin{equation}
\chi_{AB} = \sqrt{\frac{n_An_B}{m_Am_B}}k^2\left [ \omega^2 \mathbf I - \mathbf G (\kv, \omega) \right ]^{-1}_{AB},
\label{chitot}
\end{equation}
with $\mathbf I$ being the identity matrix.
\footnote{The notation in this letter is slightly different than the one in Ref.\cite{Kalman1990}. In particular $\chi_{AB}$ in Eq.~\ref{chitot} is equivalent to $\hat\chi_{AB}$ of Ref.\cite{Kalman1990}} 

The poles of $\chi_{AB}$, i.e. the zeros of the denominator of the response function, $ \mathcal D \equiv { \rm det}\left ( \omega^2 \mathbf I- \mathbf G \right )$, identify the complex collective mode frequencies of the system, which for the case of vanishing damping become the $\omega_{\pm}$ frequencies given in Ref.~\cite{Kalman2014}. With damping, the mode frequencies are shifted by a term of $\mathcal O (\bar\gamma)$. In order to relate to the observed partial fluctuation spectra we analyze the imaginary part of the respective elements of $\chi_{AB}$. Focusing on the lighter species, (species 2 in our case),
\begin{eqnarray}
\IM{\chi_{22}}&=&\frac{n_2}{m_2}\frac{k^2}{|\mathcal D|^2} \left \{ \omega R_{22}\left[ \omega^2 - \left ( C_{11} - \frac{R_{12}}{R_{22}} C_{12}\right ) \right]^2 \right .\nonumber \\
&& \left . + \, \omega R_{11} \left [ \omega^2 +  \frac{C_{12}^2}{R_{11}R_{22}}\right ] {\rm det} \mathbf R \right \}.
\label{imchi}
\end{eqnarray} 	
In the present system with the assumed dominance of the drag force in the damping, ${ \rm det} \mathbf R=0$ is satisfied and the numerator of Eq.~\ref{imchi} vanishes at 
\begin{equation}
\omega_{*1}^2 = C_{11} + \sqrt{m\rho}C_{12} = \omega_{11}^2+\Omega_{12}^2.
\end{equation}
Thus $\omega_{*1}$ becomes the location of an anti-resonance frequency for $\IM{\chi_{22}}$. Consequently $L_{22} (\omega_{*1}) $ also develops a zero, and we contend that it is $\omega_{*1}$ that appears as the sharp minimum observed in the simulation. A similar consideration for $\IM{\chi_{11}}$ provides 
\begin{equation}
\omega_{*2}^2 = C_{22} + C_{12}/\sqrt{m\rho} = \omega_{22}^2 + \Omega_{21}^2,
\end{equation}
yielding a second anti-resonance and the expectation of a minimum at $L_{11}(\omega_{*2})$. Note that the $\omega_\ast$ frequencies do not depend either on the absolute value of the damping $R_{12}$ (only the ratio of the elements of $\mathbf R$) or on the coupling constant $\Gamma$. However, the strong coupling assumption is inherent in the assumed quasi-localization of the particles. 

In Figs.~\ref{liquid} through \ref{BCC} we display the calculated plots of the pCCF for both species and compare them with the spectra obtained from MD simulation. We have chosen values of $R_{12}$ for optimal fitting; the actual values in terms of $\omega_1$ are shown in the graphs. For the liquid phase, Figs.~\ref{liquid} and \ref{gammarange}, the positions of the calculated $\omega_{*1}$ and the observed minimum in the spectrum of the lighter species $(L_{22})$ are in very good agreement. In the spectrum of the heavier species $(L_{11})$ only the appearance of a steep drop in place of the expected minimum at $\omega_{*2}$ is observed. For the higher mass ratio (Fig.~\ref{gammarange}) the agreement between the calculated and simulated spectra is much better. Note that the positions of the minima are strictly independent of $\Gamma$, as predicted by the model. At higher $\Gamma$ values, in the solid phase, the QLCA formalism in the $k\rightarrow0$ limit goes over smoothly into the conventional harmonic lattice formalism, with the understanding that the integrals over correlation functions become summation over the appropriate lattice sites. An additional condition is that the number of particles in the primitive cell of the lattice be the same as the number of species in the QLCA description. As a consequence, at $k=0$ the gap frequencies in the bcc lattice are identical to those given by the QLCA calculation. This observation allows further comparisons between the theory developed above and simulation. Comparison between MD simulation and the extended QLCA with parameters, $Z = 0.7$, $m=0.2$, $c=0.5$ is presented in Fig.~\ref{BCC}. In this case the anti-resonance is at $\omega = 1.304 \omega_1$. It is not unexpected that $R_{12}$ is much lower than in the case of the liquid. For the light component (species 2) now the agreement between theory and simulation is impressively good, not only for the location of the minimum, but for the shape of a large part of the spectrum as well. For the heavier component (species 1) the minimum is still masked, so that only a steep drop is observed in the domain of the resonances. In the high frequency domain there is no good agreement between simulation and theory and none is expected: in this domain the spectrum is governed by the details of the short time dynamics, which is not well represented by the current model.

The $\omega_{*1}$ frequency that we have discussed above can be compared with the anti-resonance of  $\RE{\chi_{22}}$. 
\begin{equation}
\RE{\chi_{22}}= \frac{n_2k^2}{m_2|\mathcal D|^2}\left \{ (\omega^2 - C_{11})\RE{\mathcal D} + \omega R_{11}\IM{\mathcal D} \right \}. 
\label{realchi} 
\end{equation}
$\IM{\mathcal D}$ is of $\mathcal O (R_{12})$, thus for small values of $R_{11}$ we can neglect the second term, $\mathcal O (R_{12}^2)$, in Eq.~\ref{realchi}. Evidently the numerator vanishes for
\begin{equation}
\omega_{01}^2 = C_{11} = \omega_{11}^2 + \Omega_{12}^2/3.
\end{equation}
These $\omega_{0A}$ frequencies are, in fact the ones associated with the normal Fano anti-resonances, referred to as Fano frequencies. They correspond to the shifted eigenmodes of the respective subsystems. The difference between the $\omega_{0A}$ and $\omega_{*A}$ frequencies is obvious, underscoring the difference between the phenomenon we are analyzing and the conventional Fano-effect. 

\begin{figure}[htb]
\onefigure[width=1\columnwidth]{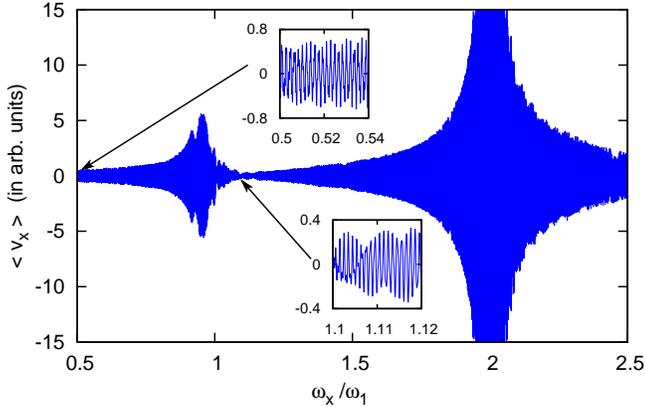}
\caption{(Color online) Amplitude of the oscillations of the longitudinal velocity of species 2 plotted against the excitation frequency of the fictitious driving force acting on the same species. The insets are magnification of the oscillations showing a decrease in amplitude at $\omega_{02} = \sqrt{C_{11}}$. $\Gamma=10\,000$, $Z=0.7$, $m=0.2$, $c=0.5$.}
\label{driving_exp}
\end{figure}

In order to see how the classical Fano effect manifests itself in the system we investigate the response of the system to an external pertubation. In the simulation a bcc lattice of 2000 point-like particles of species 1 and 2 is constructed as initial configuration and thermalized to $\Gamma=10\,000$. An external electric field acting along the $x$ direction only on species $A$ is applied in the form of $E_A (x_i,t) = \sin(\omega_x t) \sin (2\pi~x_i/ L)$. The amplitude of this field is chosen to be small (linear response regime), while the excitation frequency is linearly increased in time starting from $\omega_x = 0.4 \omega_1$ up to $\omega_x = 4 \omega_1$. During the simulation a small friction term, $-fv$, is added to the equation of motion to compensate for the heating due to the external field. The friction coefficient $f$ is continuously adjusted to provide stable temperature ($\Gamma \approx\, \rm{const}$). The time evolution of the $x$ component of the velocities is recorded to obtain $\langle v_x \rangle^B = \sum_i v_{i,x}^B \sin(2\pi~x_i/L)$ of species $B$: this measures $\RE{\chi_{BA}}$. In Fig.~\ref{driving_exp} we report $\langle v_x \rangle^2$ when only the second species is perturbed ($\RE{\chi_{22}}$). The peaks in Fig.~\ref{driving_exp} are at the zeros of $\mathcal D$ which, for the chosen parameters values, are in the vicinity of $\omega_- = 0.917\omega_1$ and $\omega_+ = 2.002\omega_1$. A significant decrease in the oscillation is evident around $\omega_{01} = 1.11 \omega_1$, which corresponds to the zero of the numerator, $\sqrt{C_{11}} = 1.1106 \omega_1$, thus confirming the consistency of the model (the zeros of $\RE{\mathcal D}$ are masked by the nearby zeros of the denominator). 

As a part of our study, a number of additional simulations have been carried out for a wide range of asymmetry parameters, $Z= \{0.5, 2\}, \, m=\{0.02,5\}, \, c=\{0.1,0.9\},$ and coupling constants $\Gamma$. All cases have been found to be in similar agreement with the theoretical predictions, based on the extended QLCA model. 

In summary we have demonstrated that the real and imaginary parts of the partial response function of a strongly coupled binary Coulomb systems vanish at characteristic frequencies of the plasma. However, the zeros of the real and imaginary parts are different. The zero of the real part of the partial response of one of the components is the resonant frequency, shifted by the coupling, of the other component, as predicted by the classical Fano effect. On the other hand, the imaginary part vanishes at a new characteristic frequency of the other component, $\omega_{*A}$. This frequency is then observable, by virtue of the FDT, in the equilibrium fluctuation spectrum. In addition, the $\omega_{*A}$ frequencies  appear to have an universal character insofar that they do not have any explicit dependence on the absolute value of the damping $R_{12}$ or on the coupling constant $\Gamma$.

While our work has addressed a strongly coupled binary coulombic system, it should be obvious from the analysis that one should expect a similar behavior in other binary or multicomponent systems, such that the system possesses more than one optic mode in its collective mode spectrum. 

A comment on the experimental measurability of our predictions is in order. It is to be realized that the partial $S_{AB}(\kv,\omega)$ spectra are, in general, not directly observable quantities (except, perhaps, in eventual microgravity complex plasma experiments). However, their linear combinations such as $S_{ZZ}(\kv,\omega)$ and $S_{MM}(\kv,\omega)$ the charge- and mass- fluctuation spectra are directly related to the scattering cross sections of electromagnetic radiation, neutron beams, etc. \cite{Hansen1979}. In these scenarios the anti-resonance pertaining to one (say 2) component may or may not be observable, depending on how much it is masked by the presence of the fluctuation of component 1. An example is provided by X-ray scattering on ``Warm Dense Matter". Here a ``dark resonance" may occur in the electronic response, as pointed out in \cite{Murillo2010}. This ``dark resonance" is, in fact the $\omega_{02}$ Fano frequency of the system. Its relationship, though, to the $\omega_{*2}$ anti-resonance identified in this work would require further clarification.
Another area of relevance may be astrophysical systems such as white or brown dwarves. The interiors of these stars consist of binary ionic mixtures with a relativistic degenerate electron gas in the background. The latter provides a very weak screening to the Coulomb interaction due to the low value of the conventional coupling constant $r_s \approx 1/137$ \cite{DeWitt1993,Jancovici1960}.

Looking ahead at the extension of the theory for finite values of $k$ or to other systems, we note that in general, the ${\rm det} \mathbf R = 0$ condition does not hold anymore since the damping matrix will contain contributions from intraspecies collision term as well. As a result, one expects the vanishing of $\IM{\chi_{AA}(\kv,\omega)}$ to occur at complex frequencies, yielding only finite minima in the fluctuation spectra.

\acknowledgments
This work has been supported by the NSF Grants PHY-0715227, PHY-0813153, PHY-1105005, OTKA Grants K-105476 and NN-103150, and the DAAD via a postdoctoral fellowship. LS would like to thank Salvatore Savo for helpful discussion.

\bibliography{References}
\bibliographystyle{eplbib}

\end{document}